\documentclass[man,floatsintext,12pt]{apa6}
\DeclareCaptionTextFormat{tabletext}{\hspace{-\parindent}\textit{#1}}
\usepackage{subfig}

\usepackage[utf8]{inputenc}    
\usepackage[T1]{fontenc}       
\DeclareUnicodeCharacter{00A0}{~}
\usepackage[american]{babel}
\usepackage{csquotes}
\usepackage[backend=biber,style=apa,sortcites=true,sorting=nyt]{biblatex}
\AtEveryBibitem{
  \clearfield{month}
}
\DeclareLanguageMapping{american}{american-apa}
\DeclareSourcemap{
	\maps[datatype=bibtex]{
		\map{
			\step[fieldsource=url,
			notmatch=\regexp{wiki},
			final=1]
		\step[fieldset=urldate, null]
		}
	}
}
\title{Re-evaluating the need for Modelling Term-Dependence in Text Classification Problems}
\shorttitle{Term-Dependence in Text Classification}

\usepackage{authblk}
\author[1]{Sounak Banerjee}
\author[2]{Prasenjit Majumder}
\author[1]{Mandar Mitra}
\affil[ ]{\textit {\{bsounak93@gmail.com, p\_majumder@daiict.ac.in, mandar@isical.ac.in\}}}
\affil[1]{CVPR Unit, Indian Statistical Institute, Kolkata, West Bengal, India; Ph : (+91) 33 2575 2858}
\affil[2]{Room No : 4209, DAIICT Gandhinagar, Gujarat, India; Ph : (+91) 79 3051 0605}

\abstract{A substantial amount of research has been carried out in developing machine learning algorithms that account for term dependence in text classification. These algorithms offer acceptable performance in most cases but they are associated with a substantial cost. They require significantly greater resources to operate. This paper argues against the justification of the higher costs of these algorithms, based on their performance in text classification problems. In order to prove the conjecture, the performance of one of the best dependence models is compared to several well established algorithms in text classification. A very specific collection of datasets have been designed, which would best reflect the disparity in the nature of text data, that are present in real world applications. The results show that even one of the best term dependence models, performs decent at best when compared to other independence models. Coupled with their substantially greater requirement for hardware resources for operation, this makes them an impractical choice for being used in real world scenarios.}

\keywords{Text Classification, Copula, Support Vector Machine, Unigram Language Model, K Nearest Neighbours}

\begin{document}
\maketitle

\section{INTRODUCTION}
For quite some time, researchers have fostered the idea of a need to design algorithms that model dependence among terms in documents to improve classification performance \cite{yu1983generalized,nallapati2002capturing,nallapati2003adaptive,eickhoff2015modelling,han2000centroid,metzler2005markov}. The central idea is that, one could better predict the class a document belongs to, if the underlying essence of the text could be interpreted, rather than an unordered collection of words that convey very little logical sense. In order to materialize the concept, many approaches have been proposed. A Copula based language model \cite{eickhoff2015modelling} presented by Eickhoff et al. considers sentential co-occurrence of term-pairs to capture the dependence structure of the terms in a document. On the other hand a centroid based document classification algorithm attempts to represent each document as a vector in the term-space and for each class, calculate a centroid vector using its constituent documents finally comparing any new documents to the available centroids \cite{han2000centroid}. While the Markov Random Field based classifier models dependence on a contiguous sequence of terms, representing them as a chained dependence structure \cite{metzler2005markov}.\\

Standard models on the other hand utilize properties such as rate of occurrence, length, distribution of features. They try to establish a relationship between the class and the properties of the documents it contains. It assigns values to features that are relevant to a particular class and then try to guess the membership of a new document to that class by comparing these values.\\

Though bolstering prediction potential through the exploitation of complex dependence structures inherent to natural language seems tempting, each of these models require significantly greater hardware resources to operate compared to independence models that as their name suggests, depend on the properties of independent features of the text. In addition to the collection of features that independence models rely on, dependence based models require both processing and memory for interpreting and storing relationships between the features.\\

In addition to that, no recent literature exists that compares classification performance of dependence models with widely accepted independence models like K nearest neighbours or support vector classifiers. So, to verify the validity of the argument justifying the use of complex dependence structures for text classification, we compare four classification algorithms which include, Naive Bayes Classifier, Copula Language Model, K Nearest Neighbour Classifier and Support Vector Machine. Each classifier is used to perform classification on multiple datasets. Finally, we analyse the merits and demerits of each classifier through a close examination of the properties of the datasets and their effects on the classifiers.\\

We used the copula based classifier as a benchmark for dependence models primarily because its superior performance over other dependence models has been well established and secondly for its recency of publication \cite{eickhoff2015modelling}.\\
 
The copula language model allows us to account for term dependence by utilizing the list of all co-occurring term-pairs in sentences. Since, each sentence in a document is the smallest entity that carries a sense, it is assumed the co-occurrence of terms in each sentence will carry some semantic relevance to the topic. The co-occurrence measures are calculated separately for all term-pairs to model a classifier for each class. The model utilizes both co-occurrence data and term probability to calculate the similarity measure of a document to a specific class.

\section{DATASETS}
Multiple Datasets were used to alleviate the possibility of any bias in the evaluation. Datasets were selected based on varying length of documents, class size, and language (colloquial and formal). Another key aspect that was considered while selecting the datasets was the classification type, multi-class and multi-label. The twitter, 20-Newsgroups and Stack Overflow datasets have been chosen for multi-class classification, while  Reuters-21578 and RCV1 are multi-label datasets.\\

All datasets were processed in the same manner. Stop word removal was carried out based on the list of English stop words available in NLTK. Stemming was done using Porter Stemmer.

\subsection{Reuters-21578}
The corpus is a collection of 21578 news wire articles from Reuters. It is a multi label dataset with a total of 90 categories. \cite{R21578} provides a detailed summary of the corpus. Class-sizes of training documents range from 1 to 2861. The average length of documents in the corpus is 126 words.

\subsection{RCV1-V2}
The original Reuters RCV1 corpus is a collection of 800,000 documents, with 103 categories. Since carrying out any operation on such a large corpus is difficult, a chronological split has been proposed \cite{lewis2004rcv1}. The RCV1-V2 contains 1 training and 4 test sets. The first 23,149 documents are used as the training set and the rest of the collection has been split into 4 test sets each containing about 200,000 documents. Each document belongs to at least 1 to a maximum of 17 categories with each topic containing atleast 5 documents over the entire corpus. For some categories there are no documents in the training set. Class-size for training documents range from 0 to 10786. Each document in the corpus is 143 words long on average.\\

Scikit-Learn provides a tokenized version of the corpus \cite{RCV1sklearn} that can be easily imported into Python. This version of the corpus was used as the input for all the existing models, except for the Copula Language Model, for which the original version that is available on request \cite{RCV1} was used, since sentence-level co-occurrence data was needed for this algorithm. For carrying out this task, documents that have been omitted from version 2 of the corpus were ignored from the original data, to maintain uniformity across all tests.

\subsection{Twitter-Sample}
This corpus has been considered because of its short document length and use of colloquial language. The corpus is available in the NLTK corpus library. It is a collection of 10,000 tweets, separated into 5,000 positive and 5,000 negative tweets. The average size of each tweet in the collection is 11 words.

\subsection{StackOverflow Questions}
Stack overflow is a platform where novices post questions from different fields and anyone who has a solution may provide an answer. The corpus contains a list of 20,000 such question titles from the stack overflow website divided over 20 categories \cite{xu2015short}. This corpus was selected because of its similarity in document size with the twitter corpus, so that any effect on classification of short text documents may be identified without bias. The average length of each question over the entire corpus is 8 words.

\subsection{20-NewsGroups}
This corpus is a collection of 18,846 news articles distributed almost evenly across 20 newsgroup categories like, \textit{comp.graphics, rec.sport.hockey, sci.electronics, soc.religion.christian}. It is available for download from their official website at \cite{20Newsgroups}, but for our purpose we used the version available in the NLTK corpus library. The class size ranges from 377 to 600 documents for training and 251 to 399 for testing. With 318 words per document, the average document size is the highest among all the datasets used in this experiment.

\section{CLASSIFICATION}

 All implementations were carried out in Python. The term-weights used for classification were kept consistent for all classifiers with the exception of the copula language model. Since the input scores of terms for copulas need to be normalized [0,1], simple probabilities of the occurrence of terms in a class were used. Where probability of occurrence of term \textit{i} in a class \textit{C} is given by,
$$P(w_i) = \frac{N_i}{\Sigma^{|C|}_{i}N_i}$$
 \(N_i\) being the number of occurrences of the term \textit{i} in class \textit{C} from the training set.\\
 
 Every other algorithm utilized the TF-IDF scores of terms for classification. The TF-IDF score of terms were generated using the \textit{TfidfVectorizer} function from Scikit-Learn.\\

 Also, in the case of RCV1 the tokenized data available from Scikit-Learn was used for our experiments. However the tokens were not labelled, so co-occurrence information could not be mapped to the original data. So, the original RCV1 data was split into sets similar to the RCV1-V2 dataset and the copula based classifier was run on this new data.\\

Finally, for performing multi-label classification on RCV1 and Reuters-21578 datasets, binary relevance method was employed. Binary relevance involves the use of a collection of yes/no classifiers, one for each class in a dataset that can determine whether a document belongs to a specific class or not.

\subsection{Naive Bayes Classifier}
It is one of the most basic and also a fairly simple classification model that uses Bayes algorithm to get probability scores. It is also the most commonly used method to benchmark other algorithms.\\
The similarity score of a document to a class is expressed as:
 $$P(t|d) = P(t)*P(d|t)$$
Where, \textit{P(t|d)} is the probability that document \textit{d} belongs to topic \textit{t}. \textit{P(t)} is the prior probability of topic \textit{t} given by:
$$P(t) = \frac{N_{t}}{N_{total}}$$
and
$$P(d|t) = \prod_{w\in d} P(w|t), $$
where \textit{\(N_{t}\)} is  the number of documents present in the training set of topic \textit{t} and \(N_{total}\) is the total number of documents present in the complete training set. \(P(w|t)\) is the probability that word \textit{w} belongs to topic \textit{t}. Though traditional Naive Bayes algorithms use simple term-probabilities, we used the TF-IDF scores of the words for the purpose of this experiment.\\

Additive smoothing was employed for smoothing of term probabilities.
The general formula for additive smoothing is

$$P(w_i) = \frac{n_i + \alpha}{N + \alpha|V|}$$

where, \(P(w_i)\) is the probability of occurrence of all words \(w_i\) in a class. \(n_i\) is the frequency of word \(w_i\) for a specific class in the training set, \textit{N} is the sum of the frequencies of all words \(w_i\) in the class, \textit{|V|} is the size of the vocabulary of the class. Finally \(\alpha\) is the user defined parameter. Laplace smoothing is a special case of additive smoothing when \(\alpha\) = 1. When 0<\(\alpha\)<1, it is called  Lidstone smoothing \cite{vatanen2010language}. For our experiments, we apply both Laplace and Lidstone smoothing (with \(\alpha\) = 0.01).\\

 We have also used the Multinomial Naive Bayes algorithm which accounts for the exact frequencies of terms for each class instead of Binomial Naive Bayes. Our reason for choosing this variation of the NB classifier is because of its superior performance in text classification problems. There have been multiple studies demonstrating the efficacy of the multinomial Naive Bayes algorithm in text classification \cite{mccallum1998comparison,eyheramendy2003naive,xu2017bayesian,protasiewicz2017categorization}.

\subsection{K-Nearest Neighbours}

 As the name suggests, for an input document \textit{d}, the KNN-algorithm selects a user-defined number of neighbours from the set of training documents, that are nearest to it.\\

 The distance between the documents is calculated using their features by plugging them into a similarity measure or graph based structures. After creating a list of \textit{K} neighbours that have the least distance, the algorithm uses a voting scheme, wherein each document enlisted by the algorithm places a vote for its respective class. The final decision for a document \textit{d} is made based on the number of votes each class receives for that document, from its \textit{K} nearest neighbours.\\

 For our experiment, the choice of \textit{K} was adjusted based on what was best suited for each corpus. A brute-force method was used to perform classification, since the implementation available could only use brute-force for sparse inputs of feature matrices. The Scikit-Learn implementation of KNN provides two options for calculating distances among documents, Euclidean and Manhattan distances. We used the Manhattan distance because the implementation of Euclidean distance caused our system to run out of memory.

\subsection{Support Vector Machine}

A support vector machine plots the documents as points in n-dimensional space where there are \textit{n} features each representing its own dimension. It then defines a hyperplane between these sets of points that segregate each set so that the collection on either side of the hyperplane contains the maximum number of documents of the intended class.\\

The function used to classify a document \textit{x} is given by:
$$ sign(\Sigma y_i*w_i*K(x`, x) + b)$$
Where:
\begin{center}
\textit{\(y_i\)} is the class value (+1 \& -1 for binary classification)\\
\textit{\(w_i\)} is the weight vector (vector for the hyperplane)\\
\textit{K} being the kernel function (linear in our case)\\
\textit{x`} is the collection of support vectors\\
\textit{b} is the distance of the hyperplane from origin\\
\end{center}

Since binary relevance was used in our case, each Support Vector Classifier (SVC) solved a binary classification problem and the sign of the value determined whether the document belonged to a class.\\

A support vector machine can use multiple functions to generate the hyperplane, these functions are called kernels. We used the linear kernel in our experiment, which basically creates a linear hyperplane. Our choice of the kernel is based on the fact that, with high dimensional vector spaces selecting non-linear kernels runs the risk of over fitting \cite{ben2010user}.

\subsection{Copula Language Model}

 In this language model, a classifier for any class \textit{c} works with two sets of features. The first one is a list of all terms present in the documents of the class with their probability of occurrence. The second is a list of all term pairs that occur in the same sentence with their respective Pointwise Multual Information (PMI) or Jaccard's coefficient values across the class, normalized between [1,\(\infty\)].
$$\theta_{t_1, t_2} = \frac{f(t_1, t_2)}{\mu}\qquad if\quad f(t_1, t_2) > \mu$$
else 
$$\theta_{t_1, t_2} = 1$$
 \textit{f} is the function for the choice of co-occurrence metric between terms \(t_1\) and \(t_2\) . \(\mu\) is the average over all \( f(t_i, t_j) \) [ \( i \neq j \) ].\\

 When theta is 1, there is complete independence between the terms and an increasing coefficient value implies increasing dependence. In our version of the algorithm we used PMI, as it generated marginally superior results compared to Jaccard in every case.\\

The probability that a document \textit{d} belongs to a topic \textit{t} is calculated by:
$$P(t|d) = P(t)*P(d|t)$$
Where \textit{P(t)} is the prior probability of topic \textit{t} and,
$$P(d|t) = C_t(w_1, w_2, w_3, . . ., w_n) $$ 
$$w_i\in d$$
Where:
$$C_t(w_1, w_2, w_3, ..., w_n) = \psi^{-1}( \psi(w_1) + \psi(w_2) + \psi(w_3) + ... + \psi(w_n) )$$ 
We used the Gumbel copula from the Archimedean family, as it was reported to produce the best results in \cite{eickhoff2014modelling}.\\

\(\psi\) and \(\psi^{-1}\) for Gumbel copulas are defined by:
$$\psi(u) = (-log(u))^\theta$$
$$\psi^{-1}(u) = exp(-u^{1/\theta})$$
Thus :
$$C_t(u_i, u_j) = exp(-( (-log(u_i))^\theta + (-log(u_j))^\theta )^{1/\theta})$$ 
Where, \(u_i, u_j\) are the probabilities of occurrence of words \(w_i, w_j\) in topic \textit{t} respectively, and \(\theta\) is a parameter that represents the strength of dependency between individual words \(w_i, w_j\) in topic \textit{t}, expressed in PMI in our case.
It is important to note that when \(\theta\) becomes 1, i.e. when they are completely independent,
$$C_t(u_i, u_j) = u_i * u_j$$ 
For the sake of simplicity, the value of \(\theta\) between two term-pairs say, \((w_i, w_j)\) and \((w_l, w_k)\) is assumed to be 1, making them independent. This causes the copula function to become a nested list of bivariate copulas.
$$C_t(w_1, w_2, w_3, . . ., w_n |c) = C_t(w_1, w_2 |c) * C_t(w_2, w_3 |c) *...$$
for all \((w_i, w_j)\) with, \(\theta\) > 1\\

We used Jelinek-Mercer smoothing for all terms in the corpus \cite{jelinek1980interpolated}. Any unfamiliar words from the test set were omitted during the classification process. Also, when considering term pairs, a single word might occur in multiple pairs, in our implementation we chose to include the contribution of the probability scores of these re-occurring terms.\\

The algorithm for this language model closely resembles the Naive Bayes algorithm. In Bayes method complete independence of terms is assumed.\\

 Given a collection of terms \(w_1, w_2, w_3, ... w_k\), from a document \textit{d} and their probabilities of occurrence \(u_1, u_2, u_3, ... u_k\) in a certain topic \textit{t}, the copula based similarity score of the document to the topic is given by:
$$C_t(u_1, u_2, u_3, ... u_k) = \psi^{-1}( \psi(u_1) + \psi(u_2)  + \psi(u_3)  + ... + \psi(u_k) )$$ 

If we assume complete independence like Naive Bayes (\(\theta=1\)),
$$C_t(u_1, u_2, u_3, ... u_k) = u_1 * u_2 * u_3 * ... * u_k$$ 
which is equal to the Bayes formula.\\

But since, this algorithm accounts for term dependence using sentential co-occurrence, it is expected to perform better than simple Naive Bayes classification.\\

The goal is to figure out whether this extra computation to improve performance using term dependence is beneficial to the text-classification paradigm.

\section{EXPERIMENTS}

 For every test except with copulas, the Scikit-Learn implementations of the algorithms were used for comparison. The input fed to each classifier was the exact same pre-processed data, generated using the standard NLTK library functions.\\

 Since there are no separate training and testing sets for the Twitter and Stack-Overflow datasets, a 10-fold cross validation was carried out for assessing the performance of each classifier. For RCV1 the classification was carried out on each test set separately. An average of the F1-measures from all 4 test sets for each classifier have been reported.\\
 
 Parameter values for almost every classifier was set to default, except for the value of K for the KNN classifier which was optimized for best result. The parameter was set to 100 for both short text datasets, StackOverflow and twitter-samples. For the rest, 15 neighbours were selected. Also the \(\alpha\) parameter for Jelinek-Mercer smoothing was set to 0.99 for the copula classifier.\\
 
The micro-averaged F1-scores of all the classifiers with the corresponding dataset have been listed in Table \ref{tab:1}.

\begin{table}
  \fontsize{10}{15}\selectfont
  \centering
  \begin{tabular}{ |c|c|c|c|c|c| } 
  \hline
  Corpus & \(NB_{\alpha = 1}\) & \(NB_{\alpha = 0.01}\) & KNN & Copula & SVC \\
  \hline
  Reuters-21578 & 0.51 & 0.77 & 0.79 & 0.64 & 0.87 \\ 
  RCV1 & 0.49 & 0.71 & 0.72 & 0.66 & 0.80\\ 
  20-NewsGroups & 0.78 & 0.84 & 0.74 & 0.82 & 0.85\\
  Stack Overflow & 0.83 & 0.78 & 0.74 & 0.76 & 0.85 \\ 
  Twitter & 0.75 & 0.72 & 0.71 & 0.70 & 0.75 \\ 
  \hline
  \end{tabular}
  \caption{Micro-averaged \(F_1\) scores.}
  \label{tab:1}
\end{table}

 \subsection{Significance Testing}

 A statistical significance test was carried out for each dataset. The Wilcoxon signed-rank test was employed for this purpose. The test was carried out by comparing the F1-scores of each classifier with the copula model. Two test strategies were used based on the type of the data.\\
 
\begin{enumerate}
	\item For the datasets that required a K-fold cross validation, the micro-averaged F1-scores for each fold were compared. In case of both, StackOverflow and Twitter corpora we used a 10-fold cross validation.
	\item For the other datasets that had a predefined train and test set, we performed a category wise testing. The F1-scores of each category were compared in order to measure whether the difference in scores were statistically significant \cite{yang1999re}.
\end{enumerate}

 Table \ref{tab:3} summarizes our observation of the significance tests. The test results have been classified into 3 categories. Category |, is for results that satisfied a confidence level of \(\alpha\)=0.01, category || were the class of results whose P-Value lied between 0.01 and 0.05 and category ||| represented a P-Value greater than 0.05. Finally, the results marked with an asterisk signify the statistical significance of the hypotheses that contradict the conclusion from Table \ref{tab:1}.

\begin{table}
  \fontsize{10}{15}\selectfont
  \centering
  \begin{tabular}{ |c|c|c|c|c| } 
  \hline
   & SVC & \(NB_{\alpha = 1}\) & \(NB_{\alpha = 0.01}\) & KNN\\
  \hline
  Reuters-21578 & | & | & ||| & |||\\ 
  RCV1 Set-1 & | & | & |* & ||*\\
  RCV1 Set-2 & | & | & |* & |||\\
  RCV1 Set-3 & | & | & |* & |||\\
  RCV1 Set-4 & | & | & |* & ||*\\ 
  20-NewsGroups & ||| & ||| & || & |\\
  Stack Overflow & | & | & | & |||\\ 
  Twitter & | & | & | & |||\\
  \hline
  \end{tabular}
  \caption{P-Value for classification performance.}
  \label{tab:3}
\end{table}

\section{RESULTS AND DISCUSSION}

 Form a rough perusal of table \ref{tab:1} we can make a few observations,
\begin{enumerate}
	\item The linear support vector classifier (SVC) outperforms the copula language model in every case in spite of the later using co-occurrence data and as a result utilizing significantly greater hardware resources.
	\item The Naive Bayes model with Laplace smoothing performs surprisingly well for short text data.
\end{enumerate}

 In order to properly analyse the results, we create a list of properties for each dataset that would enable us to explain the performance of the classifiers. These properties have been listed in Table \ref{tab:2}. The first column gives the average variance of the frequency of terms across all classes in the corpus. The stack overflow and twitter corpus have the minimum variance in term frequencies which could be a result of their short document length and the fact that they are both informal text. While the other three corpora which contain news articles that are long text data and have been constructed for formal use, have a higher variance in frequency with Reuters 21578 having the highest value. The second column contains the average document length of the corresponding corpus. And the final column of the table lists the type of classification algorithm required to perform a classification task on the corpora.\\

\begin{table}
  \fontsize{10}{15}\selectfont
  \centering
  \begin{tabular}{ |c|c|c|c| } 
  \hline
  Corpus & \(Var_{tf}\) & Doc Length & Type\\
  \hline
  Reuters-21578 & 5.228e-03 & 126 & Multi-Label\\ 
  RCV1 & 1.117e-03 & 143 & Multi-Label\\ 
  20-NewsGroups & 1.474e-04 & 318 & Multi-Class\\
  Stack Overflow & 6.845e-05 & 8 & Multi-Class\\ 
  Twitter & 1.076e-06 & 11 & Multi-Class\\ 
  \hline
  \end{tabular}
  \caption{Corpus Properties.}
  \label{tab:2}
\end{table}
 
\begin{figure}
	\begin{center}
		\includegraphics[width=0.75\textwidth]{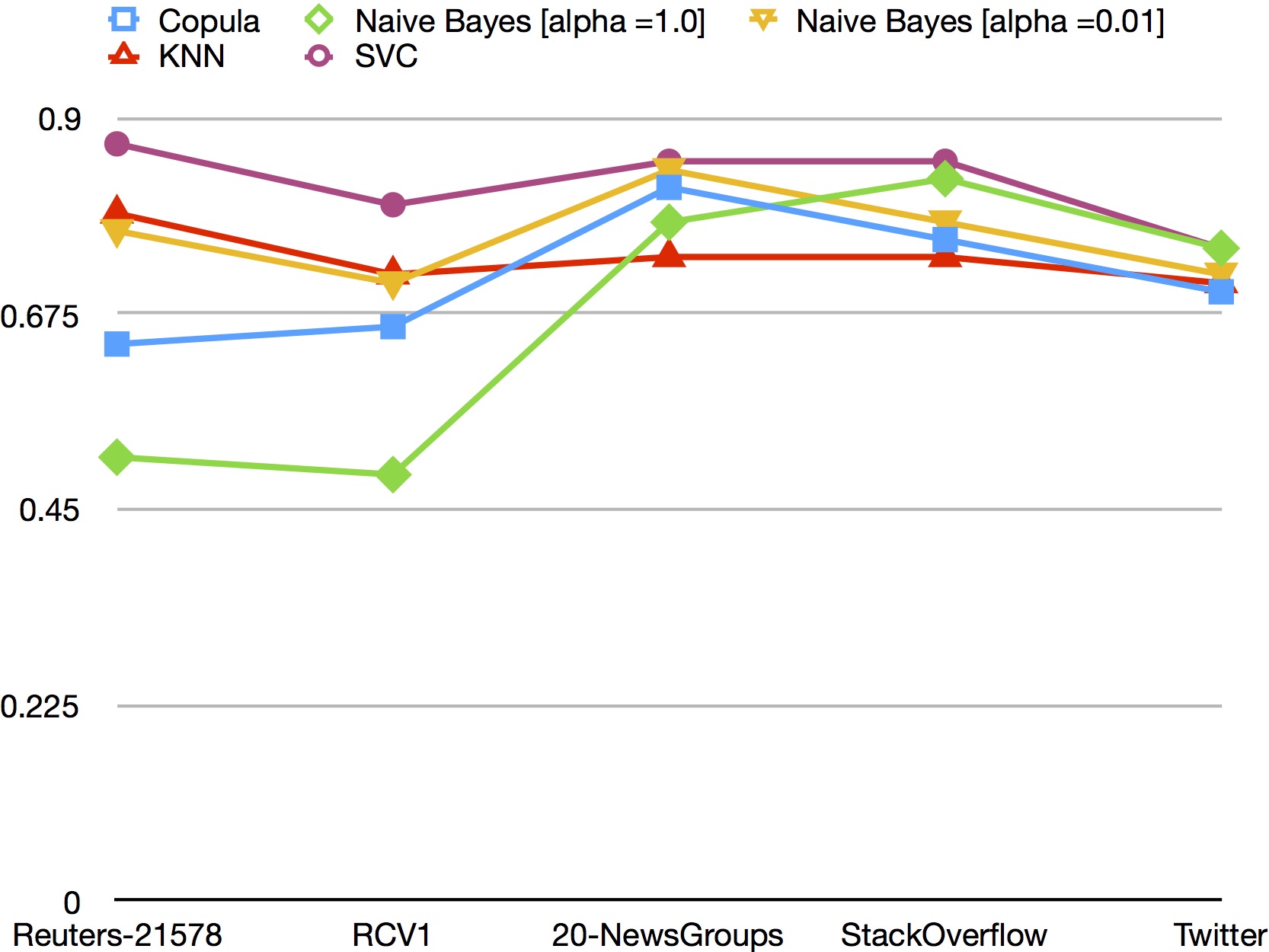}
		\caption{Plot of F1-Scores for all classifiers.}
		\label{fig:1}
	\end{center}
\end{figure}

 We studied the relationship of each of these properties against the classification scores of the models and came to the conclusion that their behaviours could be attributed to a combination of all the listed properties. But the most intuitive comparisons could be derived based on term variance. So we plot the F1-scores of the classifiers against all five corpora in decreasing order of term variance. Figure \ref{fig:1} demonstrates the the performance of each of the classifiers, copula, Naive Bayes, linear SVC and KNN respectively.\\

 On a preliminary examination of the graph, we observe that while the F1 values of both copula and Naive Bayes classifiers increases substantially when transitioning from multi-label to multi-class classification, this phenomenon does not seem to have any effect on either of the discriminative models in terms of performance.\\
 
 Both discriminative models, KNN and SVC have relatively uniform performance scores compared to the erratic nature of the scores of all the generative models. Another pattern common in almost every curve is that, datasets with higher term variance have better classification accuracy in general for both multi-class and multi-label datasets.\\
 
 Finally, from figure \ref{fig:1}, it is quite clear that the copula language model and the Naive Bayes classifier with Lidstone smoothing seem to follow a similar trend in terms of classification performance. Which proves, our earlier hypothesis about the copula language model sharing certain properties with the Naive Bayes algorithm was well founded.\\
 
 For short texts like twitter and StackOverflow both versions of the Naive Bayes classifier outperforms the copula language model. This is a result of insufficient hits on the available co-occurrence list generated from the training set, which is a direct outcome of short document length. The F1-score of KNN also plummeted when classification was carried out on short text documents, and the number of neighbours had to be adjusted to 100 to improve accuracy.\\
 
 Moving on to Table \ref{tab:3} we observe that, even with a substantial difference of micro-averaged F1-scores of copulas and Naive Bayes with Lidstone smoothing for the Reuters-21578 data it has a confidence score greater than 0.05. More surprisingly the significance score of this algorithm for the RCV1 corpus indicates, with a 0.01 confidence level that copula is a better algorithm. KNN presented similar counter-intuitive significance scores for both multi label corpora, with a very high margin of difference in their micro-averaged F1-scores. SVC and Naive Bayes with Laplace smoothing also have P-values higher than 0.05 for the 20-NewsGroups dataset.\\
 
 Similar results were observed for the Twitter and StackOverflow corpora in case of KNN, but the differences in its micro-averaged F1-scores with copulas were very low in both cases and no relevant conclusion could be drawn even with a more detailed analysis. So, the observations were attributed to data bias.\\
 
\begin{figure}
	\begin{minipage}{0.4\textwidth}
		\includegraphics[width=\textwidth]{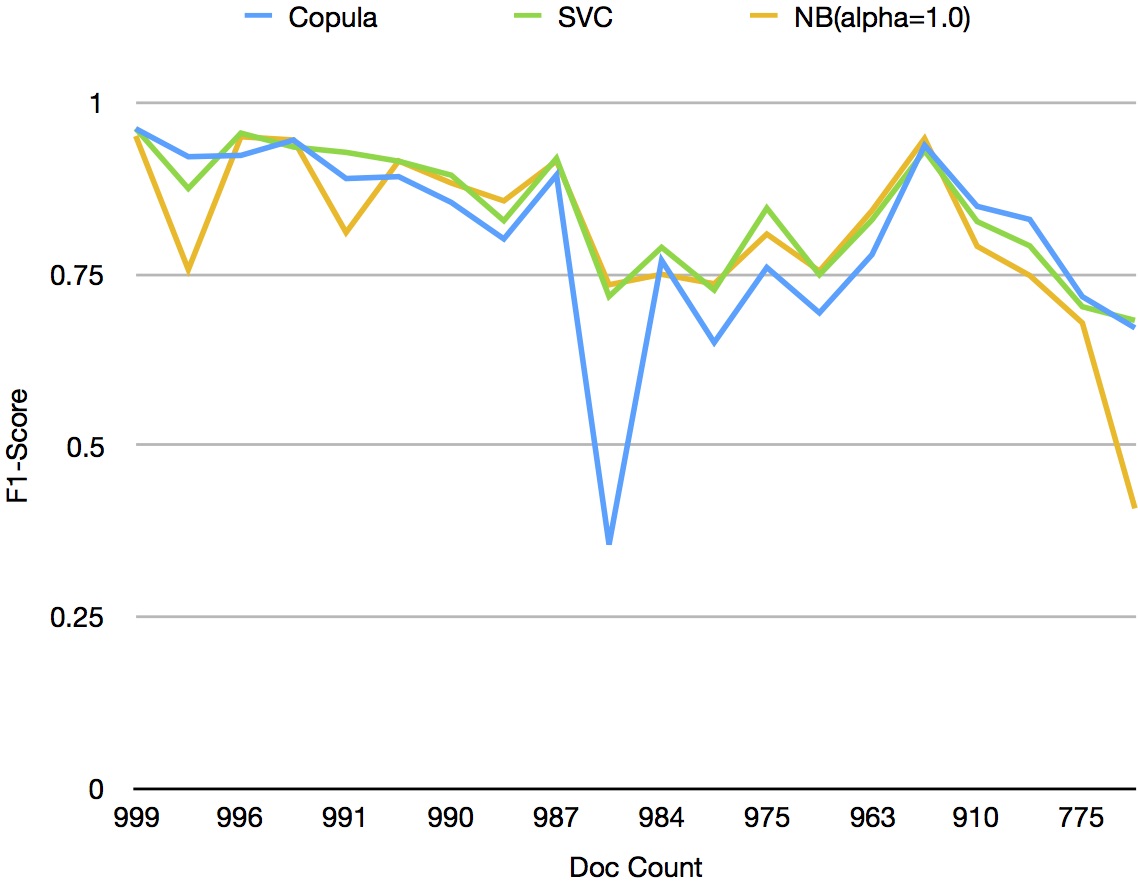}
		\caption{Plot for F1-Score across all classes for 20-NewsGroup Corpus}
		\label{fig:5}
	\end{minipage}
	\hfill
	\begin{minipage}{0.4\textwidth}
		\includegraphics[width=\textwidth]{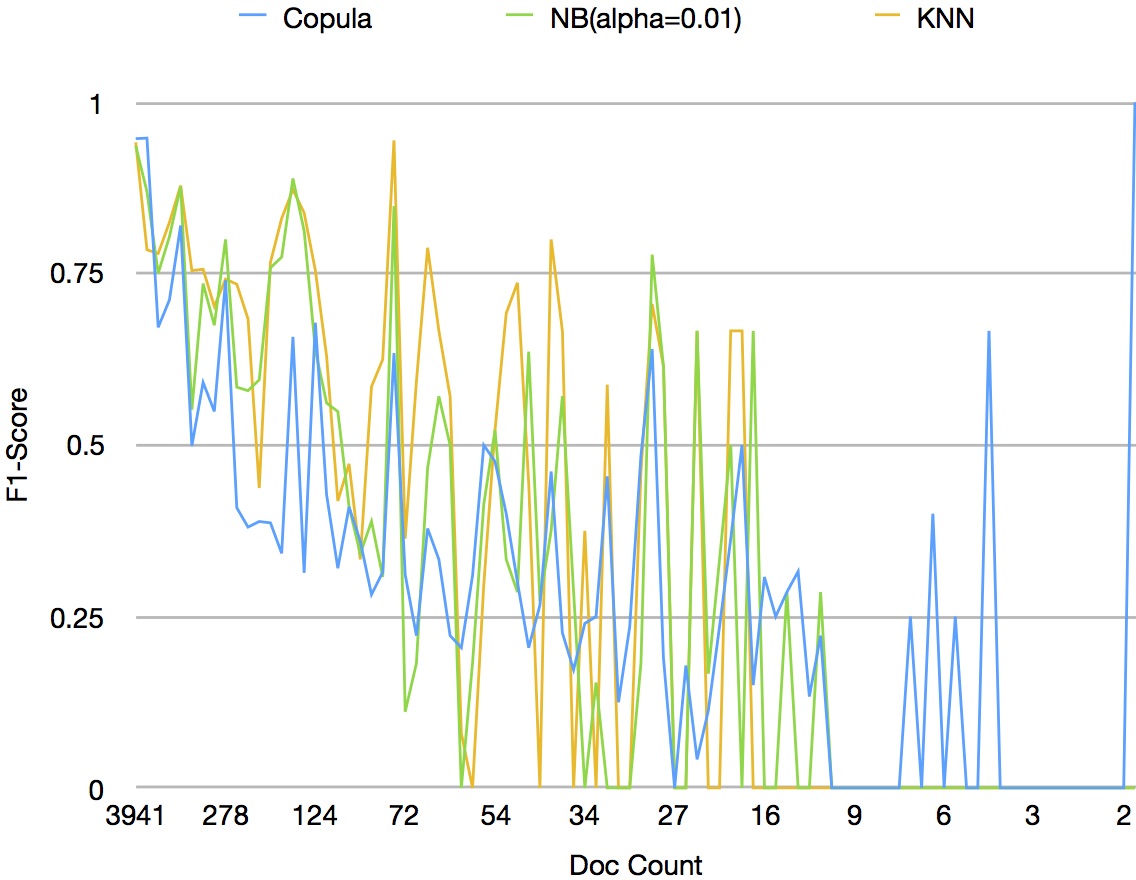}
		\caption{Plot for F1-Score across all classes for Reuters-21578 Corpus}
		\label{fig:6}
	\end{minipage}
	\\
	\\
	\begin{minipage}{0.4\textwidth}
		\includegraphics[width=\textwidth]{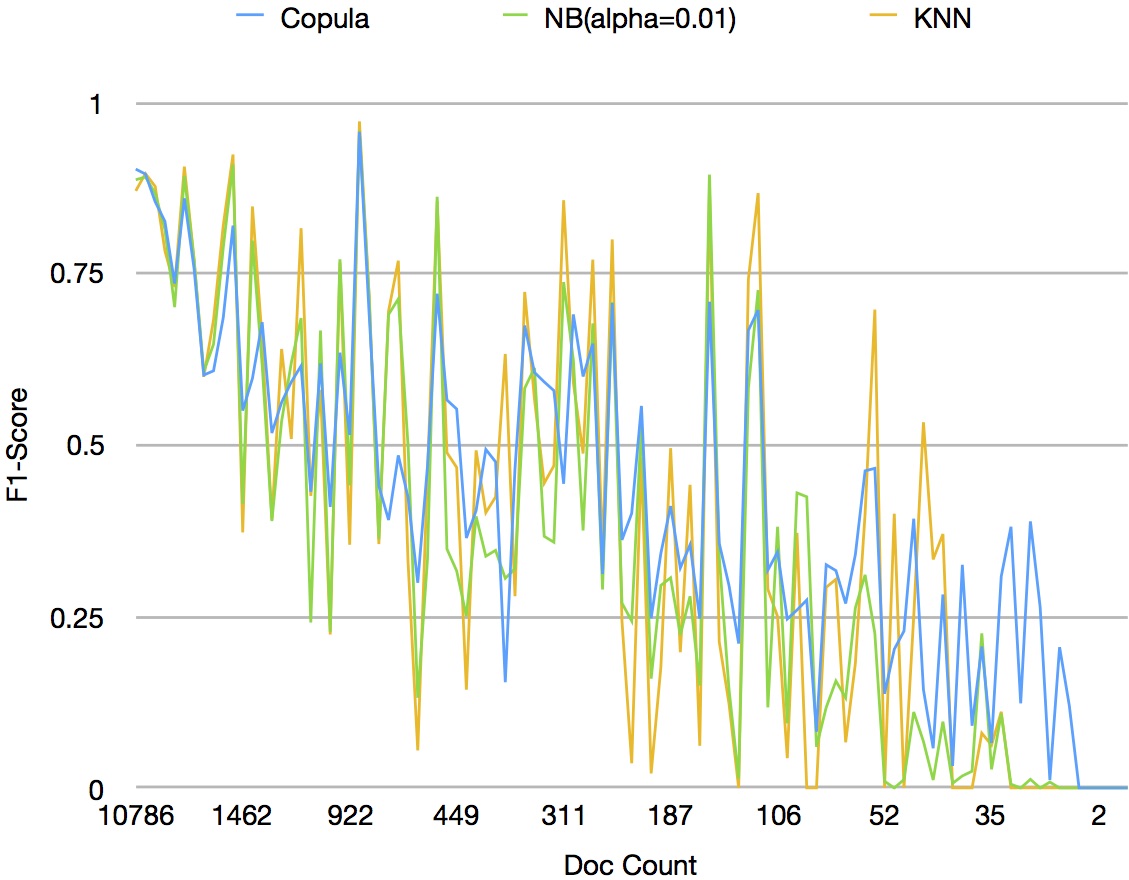}
		\caption{Plot for F1-Score across all classes for test set-1 of RCV1 Corpus}
		\label{fig:7}
	\end{minipage}
	\hfill
	\begin{minipage}{0.4\textwidth}
		\includegraphics[width=\textwidth]{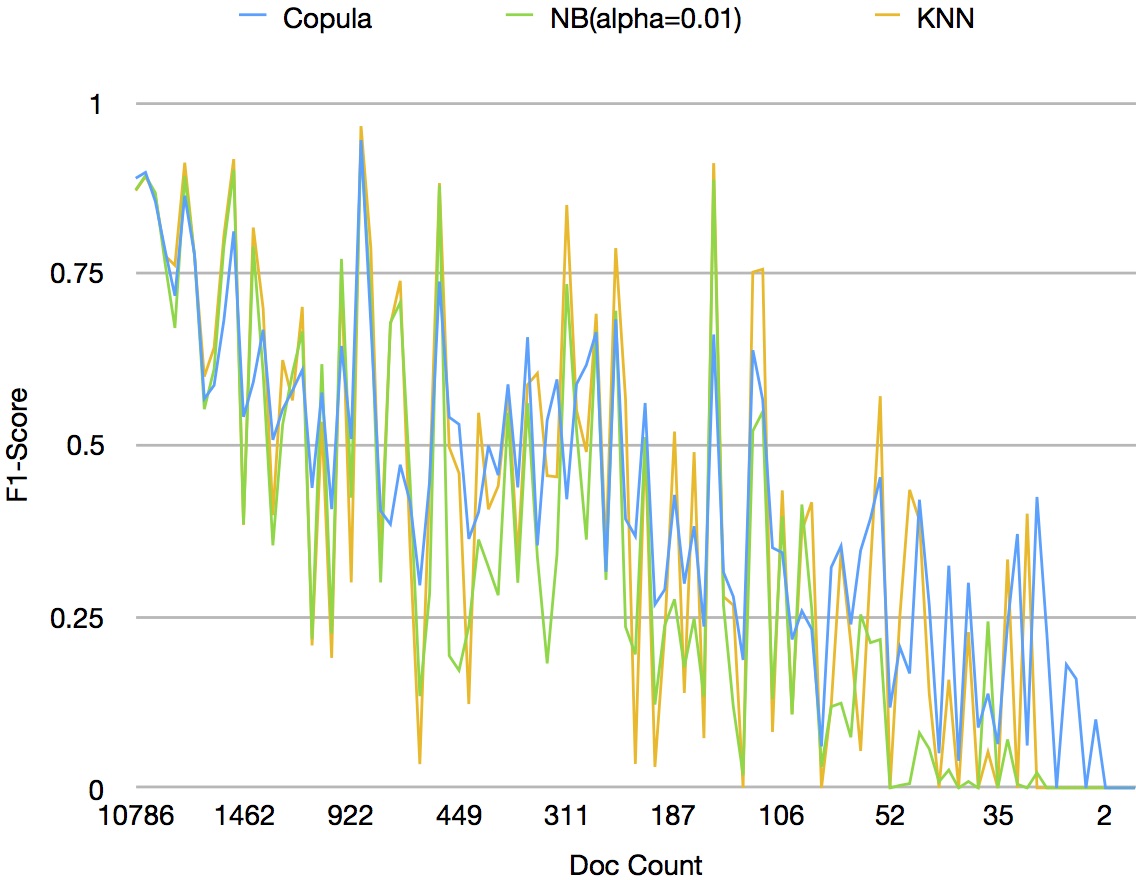}
		\caption{Plot for F1-Score across all classes for test set-2 of RCV1 Corpus}
		\label{fig:8}
	\end{minipage}
	\\
	\\
	\begin{minipage}{0.4\textwidth}
		\includegraphics[width=\textwidth]{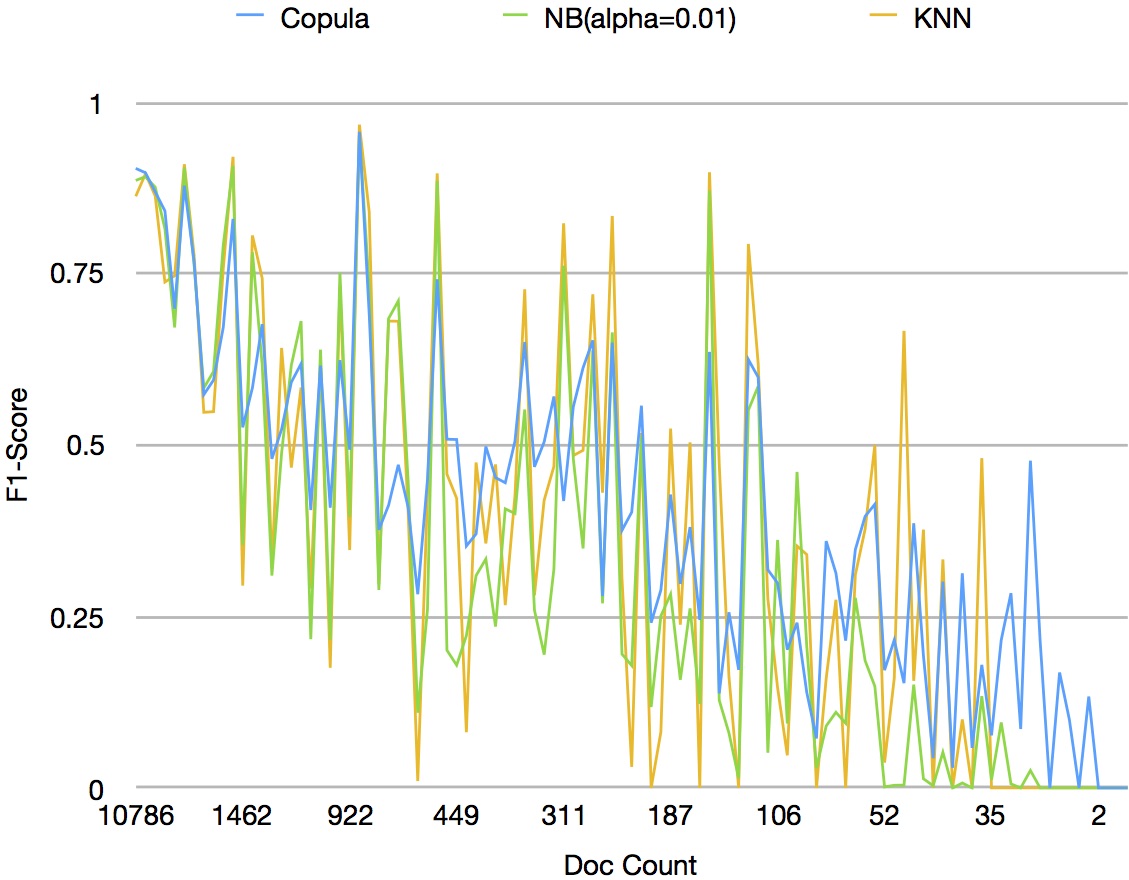}
		\caption{Plot for F1-Score across all classes for test set-3 of RCV1 Corpus}
		\label{fig:9}
	\end{minipage}
	\hfill
	\begin{minipage}{0.4\textwidth}
		\includegraphics[width=\textwidth]{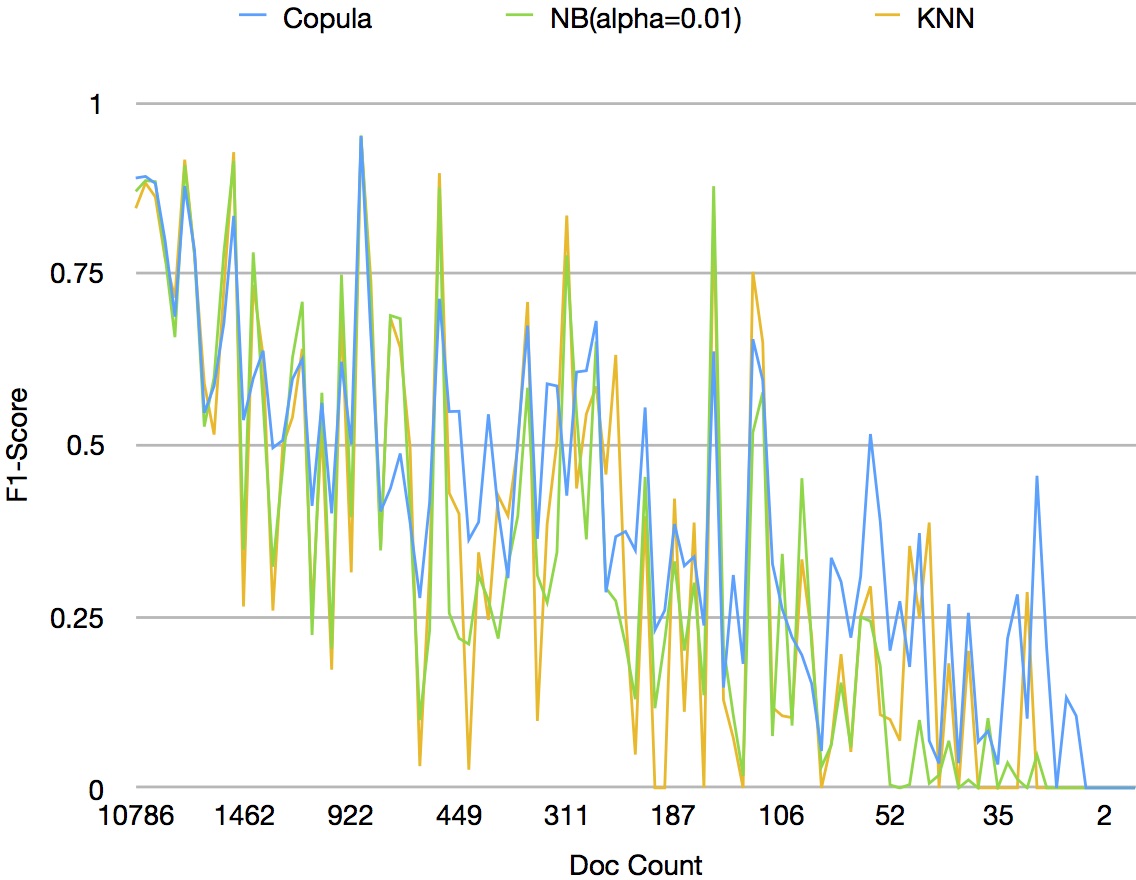}
		\caption{Plot for F1-Score across all classes for test set-4 of RCV1 Corpus}
		\label{fig:10}
	\end{minipage}
\end{figure}

 In order to create a better understanding of the anomalies in the confidence scores for the rest, we generated Charts \ref{fig:5} through \ref{fig:10} that demonstrate the performance scores of each classifier over all the data points, that were used to perform the significance tests. The X-axis lists all classes in a corpus in decreasing order of class-size and the Y-axis plots the F1-scores.\\

 For SVC in figure \ref{fig:5} the differences in scores are not as significant, but there are 6 cases where copula marginally outperforms this model. This resulted in a marginally higher P-value.\\
 
 But more interestingly, in each case the copula model demonstrates a significant and consistent improvement in classification accuracy over other algorithms, for classes that have a low document frequency. Thus the superior performance of the classifier in such categories is the cause for the shift in the significance values. This also sheds new light on the properties of the copula classification model. We learn that the information the model accumulates using term dependence helps with classification accuracy for classes with inadequate features.\\
 
\begin{figure}
	\begin{center}
		\includegraphics[width=0.75\textwidth]{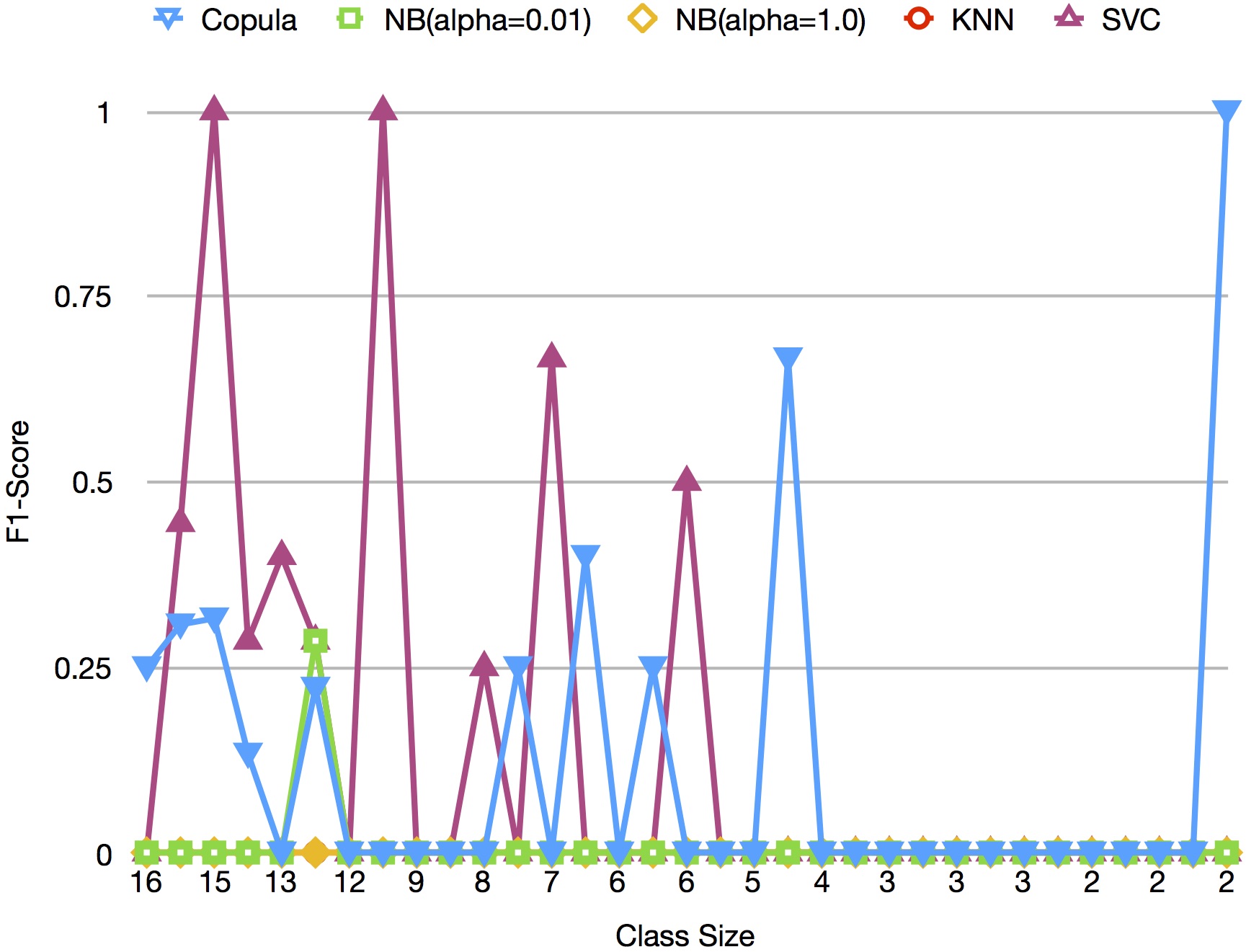}
		\caption{Plot of F1-Scores of all classifiers for small classes from the Reuters-21578 Corpus.}
		\label{fig:Mag_R90}
	\end{center}
\end{figure}

To further investigate this property, we decided to plot the F1-scores of all classifiers for only the classes that had the highest difference in performance. By studying graphs \ref{fig:5}-\ref{fig:10}, we observe that this phenomenon is most clearly visible for class sizes 16 through 2 of the Reuters-21578 corpus. Figure \ref{fig:Mag_R90} presents the accuracy measures of all the classifiers for the aforementioned sequence of classes. The superior performance of the dependence model is clearly visible in the chart.\\

\begin{figure}
	\begin{minipage}[b]{0.4\textwidth}
		\includegraphics[width=\textwidth]{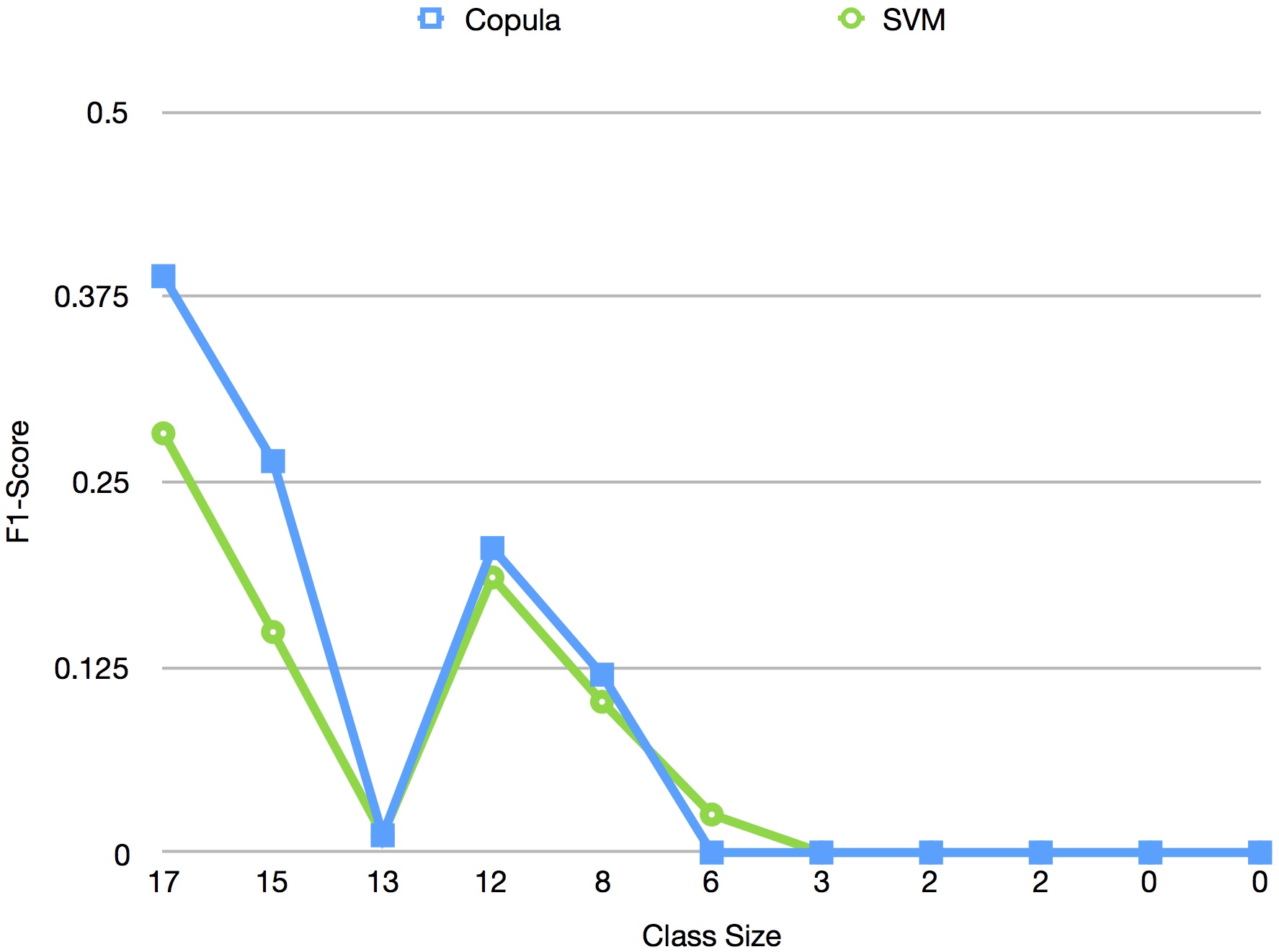}
		\caption{F1-Scores for small classes from RCV1, Set 1}
		\label{fig:Mag_RCV1S1}
	\end{minipage}
	\hfill
	\begin{minipage}[b]{0.4\textwidth}
		\includegraphics[width=\textwidth]{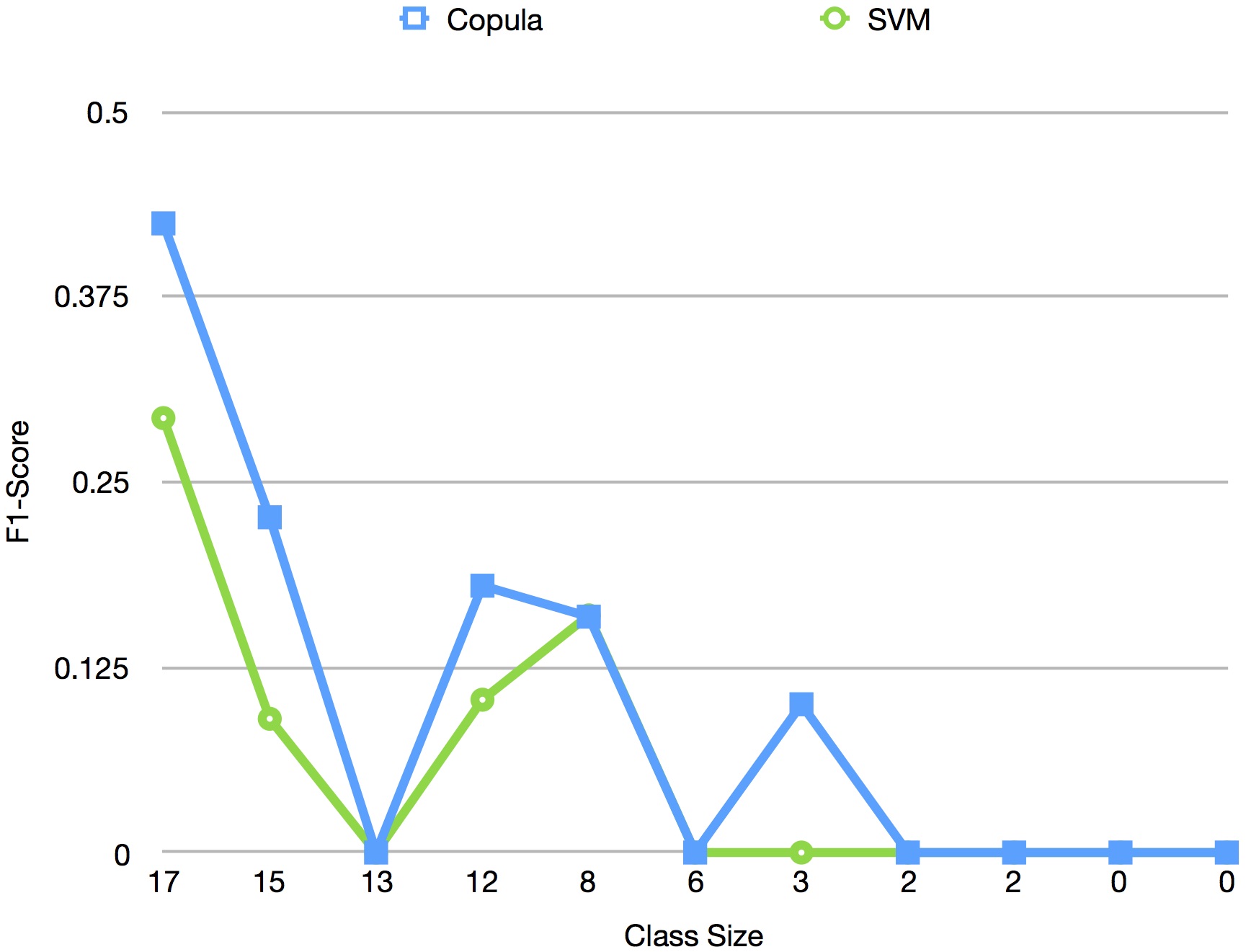}
		\caption{F1-Scores for small classes from RCV1, Set 2}
		\label{fig:Mag_RCV1S2}
	\end{minipage}
	\\
	\\
	\begin{minipage}[b]{0.4\textwidth}
		\includegraphics[width=\textwidth]{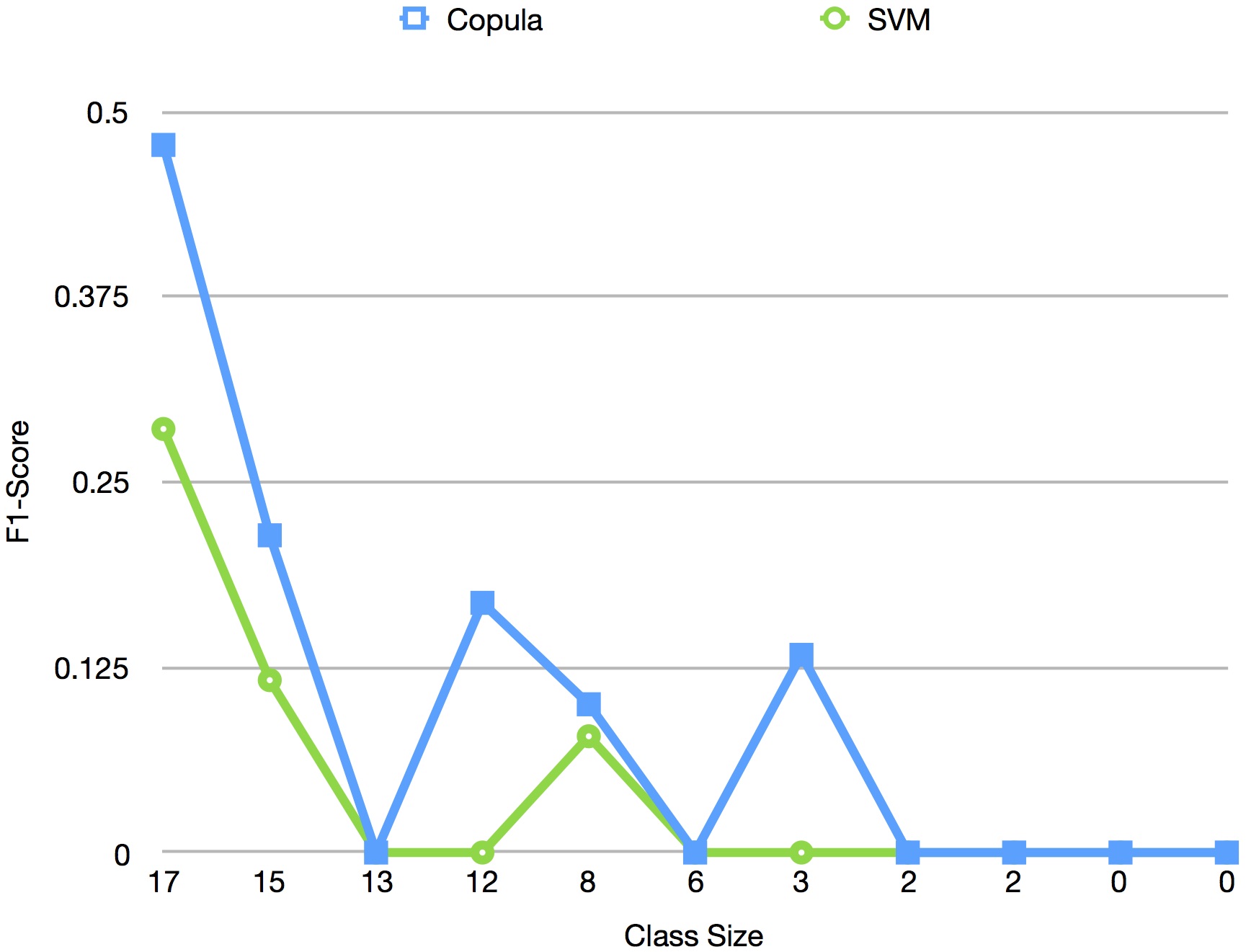}
		\caption{F1-Scores for small classes from RCV1, Set 3}
		\label{fig:Mag_RCV1S3}
	\end{minipage}
	\hfill
	\begin{minipage}[b]{0.4\textwidth}
		\includegraphics[width=\textwidth]{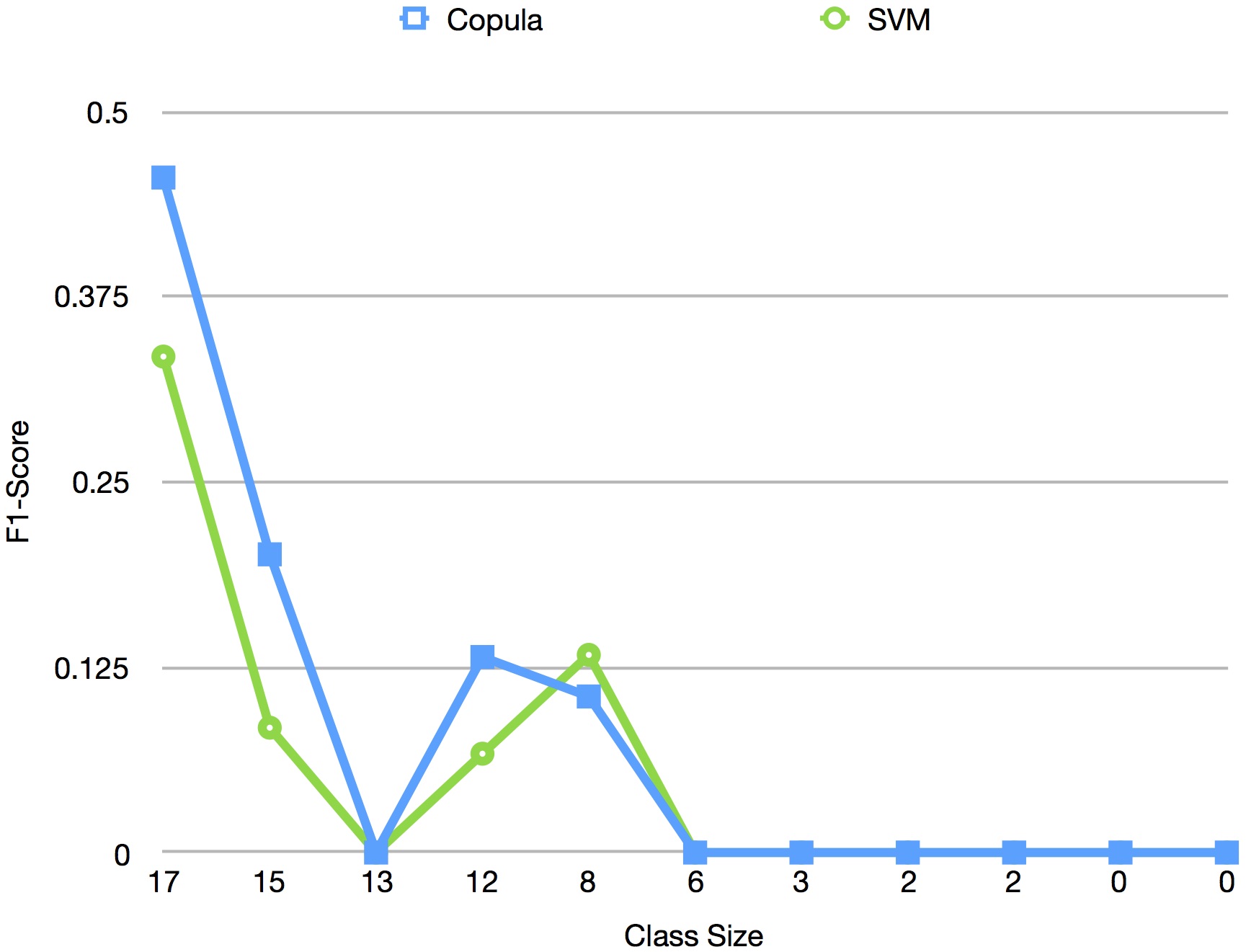}
		\caption{F1-Scores for small classes from RCV1, Set 4}
		\label{fig:Mag_RCV1S4}
	\end{minipage}
\end{figure}

 But, even with copulas showing impressive performance, SVC still manages to do a better job at classifying documents in most cases. To eliminate the possibility of a bias, we plot the F1-scores of the two classifiers for a similar range of class size from a different corpus. The RCV1 corpus was the only other corpus with comparable class sizes, so we used the results of the four test sets to compare the performance of the two classifiers. Figures \ref{fig:Mag_RCV1S1} to \ref{fig:Mag_RCV1S4} represent the F1-scoes of the smallest classes of each test set for the two classifiers. In all 4 cases copulas clearly take the lead.\\

 Thus it can be concluded that copulas evenly match SVM based classification in classes with sparse features, for long text data. The relative scores of the two classifiers will depend on the nature of data, but in general both algorithms manage to perform decently for such classes.

 \section{CONCLUSION}

 From the extensive set of experiments that were carried out, it is clear that Support Vector Machine based classifiers continue to dominate over others and remain the most reliable classifier. All the other classifiers had their own limitations. Even though the copula model demonstrated impressive performance for classes with a limited number of documents, SVM achieved nearly equal performance in fractional time. It also performed poorly in short text datasets, where Naive Bayes demonstrated why it still remains a benchmark for all other classification algorithms. While classification accuracy for generative classifiers faltered in multi-label classification problems compared to multi-class, discriminative methods maintained a very stable curve. But most importantly, the copula language model in spite of boasting the use of complex dependence structures, failed to impress.\\
 
 It is therefore clear that dependence models like the Copulas are still outperformed by common independence methods like KNN and SVM and even small modifications to the Naive Bayes Classifier like changing smoothing parameters can sometimes result in better scores. \\
 
 A state of the art dependence model could not hold its place among existing classification algorithms that do not model term dependence, which also makes them less resource intensive than the former. So the question remains, should researchers continue to introduce new dependence models that perform better than their predecessors or focus on improving the performance of existing methods?\\
 
 The inherent limitation of modelling term dependence on text data lies in the considerably high costs associated with computation and storage, which in turn creates a demand for a significantly higher classification accuracy from these algorithms. Term dependence obviously has its perks, but every algorithm that models this dependence can not automatically be expected to be better than existing state of the art models like SVM. Proposing dependence models that perform gracefully when compared to the best existing models even for specific use cases could be a valuable contribution. But, introducing new algorithms that generate marginally better results than models which are themselves not very efficient, may not be in the best interest for the progress of the field.

 \section{ACKNOWLEDGEMENT}
 
 We would Like to thank Kaggle and Scikit-Learn  for for making the StackOverflow and a tokenized version of the RCV1-V2 dataset available to us for our experiments.

\printbibliography

\end{document}